\documentclass[twoside]{article}
\usepackage[latin1]{inputenc}
\usepackage{t1enc}
\usepackage{a4}
\usepackage{tabularx}
\usepackage{epsf}
\usepackage{psfig}

\def\bref{\vspace{4pt}\noindent\hangindent=10mm}
\def\degr{\hbox{$^\circ$}}
\def \la{\mathrel{\mathchoice   {\vcenter{\offinterlineskip\halign{\hfil
$\displaystyle##$\hfil\cr<\cr\sim\cr}}}
{\vcenter{\offinterlineskip\halign{\hfil$\textstyle##$\hfil\cr
<\cr\sim\cr}}}
{\vcenter{\offinterlineskip\halign{\hfil$\scriptstyle##$\hfil\cr
<\cr\sim\cr}}}
{\vcenter{\offinterlineskip\halign{\hfil$\scriptscriptstyle##$\hfil\cr
<\cr\sim\cr}}}}}

\begin{document}

\setcounter{figure}{0}
\setcounter{section}{0}
\setcounter{equation}{0}

\begin{center}
{\Large\bf
A Map of the Northern Sky:\\[0.2cm]
The Sloan Digital Sky Survey in Its First Year}\\[0.7cm]

Eva K.\ Grebel \\[0.17cm]
Max Planck Institute for Astronomy\\
K\"onigstuhl 17, D-69117 Heidelberg, 
Germany\\
grebel@mpia-hd.mpg.de
\end{center}

\vspace{0.5cm}

\begin{abstract}
\noindent{\it
The Sloan Digital Sky Survey (SDSS) is the largest and most ambitious
optical CCD survey undertaken to date.  It will ultimately map
out one quarter of the sky with precision photometry in five
bands, high-quality astrometry, and spectra of all galaxies and
quasars brighter than certain limiting magnitudes.  
The scientific potential of the SDSS
is enormous and addresses
a wide variety of astrophysical key questions. 
After a proprietary period the reduced and
calibrated data are made available to the astronomical community as a whole.

The SDSS is run 
by an international consortium involving universities and research
institutions.  One of the participating partners is
the Max Planck Institute
for Astronomy in Heidelberg.  
Already in its first year the SDSS has led to a number of spectacular
scientific discoveries.  In this paper, we will introduce the SDSS
survey, discuss its scientific potential, highlight important science
results, and present the Calar Alto key program for SDSS follow-up studies. 
}
\end{abstract}

\section{Introduction}
Sky surveys have a tradition dating back many centuries.  Early catalogs
and charts were based on eye estimates of positions and luminosities, which
gained considerably in depth and accuracy with the introduction of telescopes
and measurement tools such as the meridian circle.  The invention of 
photographic plates enabled astronomers to record large areas of the sky
efficiently and objectively.  Digitized versions of optical photographic 
all-sky surveys are presently widely used and are often available on-line,
accompanied by catalogs with detailed photometric and astrometric information
both for point sources and extended objects.  Spectroscopic follow-up
surveys resulted in spectral classification of bright stars, proper motions,
or classification and redshifts of luminous galaxies.  The optical 
surveys are complemented
by all-sky surveys in other wavelength ranges, which continue to gain 
in sensitivity and resolution.

The Sloan Digital Sky Survey (SDSS) is a large new optical survey that is purely
CCD-based.  The SDSS is expected to ultimately cover up to 10,000
square degrees 
centered on the north Galactic cap ($\alpha = 12^{\rm h}20^{\rm m}$,
$\delta = +32\degr30'$) and three great circle slices of
a total of 225 square degrees near the south Galactic cap
($\alpha = 20.7^{\rm h}$ to $4^{\rm h}$ at $\delta = 0\degr$; 
$\alpha = 20.7^{\rm h}$ to $22.4^{\rm h}$ at $\delta = -5.8\degr$;
$\alpha = 22.4^{\rm h}$, $\delta = 8.7\degr$ to $\alpha = 2.3^{\rm h}$,
$\delta = 13.2\degr$).  The total area corresponds to
one quarter ($\pi$ steradians)
of the sky.  The scan regions were selected to avoid areas of high
Galactic extinction.  The SDSS provides 5-filter photometry,
astrometry, and spectroscopy.  
It is estimated that the imaging survey will ultimately comprise 
$\sim 8 \cdot 10^7$ stars, $\sim 5 \cdot 10^7$ galaxies, and $\sim 10^6$
quasars with high-quality photometry.  

The participating institutions in the SDSS collaboration and some of
the rules under which the collaboration operates are presented in Section 2.
Technical details on the SDSS and a summary of its data products
are given in Section 3.

The SDSS camera saw first light in June 1998.  After the commissioning phase
regular operations began in April 2000.
Calibrated SDSS data will be made publicly available after a proprietary
period,
benefiting the astronomical community as a whole.  The first
incremental data release is scheduled for June 2001. 
The SDSS project will continue to take data over a total of five years.

The SDSS is unprecedented
in its combination of depth, homogeneity, and area coverage.  
Its scientific potential is enormous and allows one to 
address a wide variety of astrophysical key questions.  
It provides an
invaluable database for areas such as
local star formation, stellar and galaxy luminosity functions, Galactic
structure, low-surface brightness galaxies, galaxy evolution, clusters
of galaxies, lensing, large-scale structure, and cosmology. 

In this paper we attempt to  summarize important scientific findings derived
from SDSS data so far (Section 4)
and present the Calar Alto key project for SDSS
follow-up observations (Section 5). 

\section{Survey Participants and Operations}

The SDSS is an international project with meanwhile 
11 participating institutions in the USA, Japan, and Germany.  
The current institutional
participants are the Fermi National Accelerator Laboratory, the
Institute for Advanced Study, the Japan Participation Group, the
Johns Hopkins University, the Max Planck Institute for Astronomy (MPIA),
the Max Planck Institute for Astrophysics (MPA), New Mexico State University,
Princeton University, the University of Chicago, the 
United States Naval Observatory, and the University of Washington.   
About 200 astronomers (including students) are involved in various aspects
of SDSS research.

A collaboration of this size requires a well-defined set of rules and
responsibilities in order to function.  Data access and usage 
are subject to a mutually agreed upon set of guidelines.  Planned 
science projects are announced within the collaboration to give all
interested parties an opportunity to join and to avoid conflicts of
interest.  Publications
are reviewed within the collaboration before submission.  
At the US universities that originally started the SDSS every faculty or staff
member and their students and postdocs may work with SDSS data if 
they desire for as long as they work at these SDSS institutions.  At 
institutions that joined later the number of participants
is limited by memoranda of understanding.
There is a group of so-called external participants who are not at SDSS
institutions but who were awarded
data rights for specific projects.  
Finally there are ``builders,'' a term that
comprises both scientific and technical personnel that contributed 
significantly to the survey in its initial phase.  Builders retain data 
rights even when leaving
SDSS institutions and may request co-authorship on any SDSS publication 
they wish.  

MPIA joined the SDSS in 1999.  Eight people (including two participants
as well as students and postdocs) are actively working on 
SDSS projects.  As an in-kind contribution to the SDSS, MPIA has begun a
five-year key project for SDSS follow-up studies at Calar Alto (Section 5).  
MPA joined the project in 2001.  Other German participants include
external collaborators at the Max Planck Institute for Extraterrestrial
Physics, who contribute through the ROSAT All-Sky Survey (RASS).

\section{SDSS Technical Information}

The SDSS uses a dedicated 2.5-m telescope at Apache Point Observatory
($l = 32\degr 46' 49.30''$ N, $b = 105\degr 49' 13.50''$ W, 
elevation 2788 m) in New Mexico, USA.  The telescope is a 
f/5 modified Ritchey-Chr\'etien altitude-azimuth design.
It allows distortion-free imaging over a $3\degr$ wide field through a 
large secondary mirror and two corrector lenses.

\subsection{The Photometric Survey}

The imaging part of the SDSS is carried out as a
simultaneous five-color optical drift-scan survey.   
Drift scanning eliminates overhead due to read-out times and 
pointing offsets and minimizes adverse effects of pixel-to-pixel
variations.  It provides an efficient observing mode with superior
flat fielding and image uniformity.  The multi-filter observations
reduce effects of sky background variations and ensure that all
data of a field are taken at the same airmass.  On the other hand,
drift scanning is limited in depth by telescope aperture and
instrument sensitivity.  

The five SDSS filters are a modified
Thuan-Gunn system ($u', g', r', i', z'$; Fig.\ 1).  They were 
designed to provide
a wide color baseline with minimum overlap, to avoid night sky lines and 
atmospheric OH bands,
to match passbands of photographic surveys, and to guarantee good
transformability to existing extragalactic studies.
The SDSS
photometric system is described in Fukugita et al.\ (1996) and in
Lupton et al.\ (1999).  

\begin{figure}[ht]    
\centerline{\vbox{
\psfig{figure=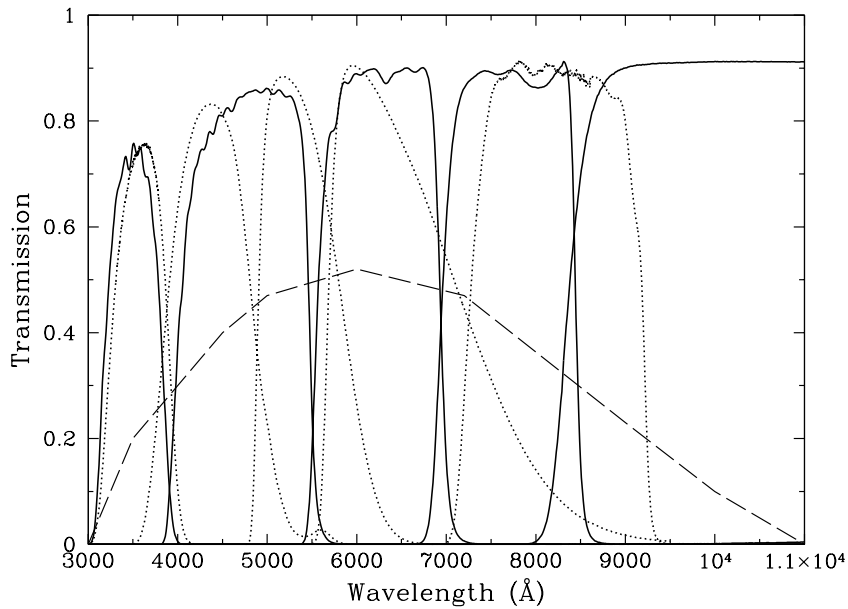,width=12cm,angle=270}
}}
\vspace{-0.3cm}
\caption{ \label{fig_filters}
{
SDSS filter transmission curves (solid lines, from blue to red: $u', g', r',
i', z'$) compared to Johnson-Cousins filter curves (dotted lines,
from blue to red: {\it U, B, V, R, I}).  The dashed line indicates the
SDSS system response.  
}
}
\vspace{-0.4cm}
\end{figure}

The SDSS imaging 
camera consists of a rectangular array of 30 Tektronics/SITe
CCDs arranged in 5 rows $\times$ 6 columns.  The CCDs have $2048 \times 2048$
pixels each.  The pixel size is $24 \mu m$, providing a pixel scale
of $0.4''$.  Typical seeing values are $1.5''$, but efforts are undertaken
to improve the seeing.  Each CCD provides a field of view with a width of
$13.6'$.  The gap between CCD columns is $12'$.  
Each scan (referred to as strip)
covers a non-contiguous area with a width of $\sim 1.36\degr$.
To fill the gaps a second scan is made, offset such that the overlap with 
the previous scan lines is $50''$ on either side.  The width of completed
scan (referred to as stripe) is thus $\sim 2.5\degr$. 
The engineering and technical details of the SDSS photometric camera  
are given in Gunn et al.\ (1998).

Each column of CCDs was optimized
in sensitivity for its assigned filter passband.  Each column is preceded
and succeeded by rows of 22 smaller CCDs ($2048 \times 400$ pixels)
used for astrometric calibration.  SDSS
astrometry is based largely on the Hipparcos and Tycho catalogs
and is accurate to $\sim 50$ mas.

When used at sidereal scanning rate the effective exposure time is 54 s,
resulting in an $r'$-band magnitude of $\sim 22.3$ mag at S/N = 10
(for point sources).  In the
southern survey repeated imaging of the stripes is planned to allow
one to search for variable objects and to reach about two magnitudes fainter.

The photometric calibration of the survey is provided by a 50-cm 
telescope on site (the Photometric Telescope), which measures atmospheric
extinction, sky brightness, and photometric standards.  

For further details, refer to the on-line SDSS Project book at\\
{\centerline{\tt http://www.astro.princeton.edu/PBOOK/welcome.htm}.}  
For a technical summary of the SDSS, 
see York et al.\ (2000). 

\subsection{The Spectroscopic Survey}

The SDSS obtains
 medium-resolution spectroscopy covering the same area as the photometric
survey.  The spectra are taken with a multi-object fibre spectrograph with a 
field of view of $3\degr$.  There are 640 fibers in total, out of which 
$\sim 480$ are used for galaxies, $\sim 100$ for quasars,
$\sim 20$ for other targets of special interest, $\sim 32$ for 
sky spectra, and the remainder for standards.
The fibre plug plates are custom-drilled for each 
field, and the fibers are inserted by hand.
Each fibre has a diameter of $3''$ (0.2 mm).  The minimum fibre 
separation is $55''$.  In order to account for variable galaxy density 
the overlap between plug plates is selected such that the spectroscopic
coverage is optimized (adaptive tiling).

The SDSS spectrograph is a two-channel spectrograph covering a wavelength
range of 3900 \AA\ to 9100 \AA\ at a resolution of $\sim 3$ \AA\ 
($1800 < R < 2100$).  Half of the 640 fibers are fed into the 
blue (3900 \AA\ --  6200 \AA ) spectrograph, and the other half into
the red (5800 \AA\ -- 9200 \AA ) spectrograph.
The
velocity resolution is $\sim 150$ km s$^{-1}$ (70 km s$^{-1}$ pixel$^{-1}$)
with a 10 -- 15 km s$^{-1}$ radial velocity accuracy.  Exposure times are
typically 45 min ($3 \times 15$ min).  For faint sources,
S/N $> 13$ per \AA . 

The targets are selected from the photometric survey.  Galaxies and 
quasars receive priority over stars as reflected in the number of 
fibre assignments listed above.  SDSS is expected to ultimately obtain 
spectra of 
900,000 field galaxies ($r'\la 18.2$ mag, median redshift 0.1),
100,000 luminous red galaxies volume-limited to $z \sim 0.4$, and 
100,000 quasar candidates ($g' \la 19.2$ mag).  
Stars are only targeted when they have unusual colors. 
Further details can be found in the SDSS Project Book available on-line
at the URL given in Section 3.2.

\subsection{Data Products}

Imaging data and spectra are reduced with automated pipeline packages
written specifically for SDSS data.  

The imaging survey yields both
corrected images and photometry tables.  The images are 
useful for, e.g., the search for low-surface-brightness features
or the production of finder charts.  Cut-out pixel
masks, so-called atlas images, are available for each detected object
in all passbands.  For a quick look, combined multi-color gif images
are available for each scan.  

The photometry tables contain all point sources and extended sources
detected in the images, their positions, fluxes and magnitudes in the 
five filters, Galactic foreground reddening estimated from the maps by
Schlegel et al.\ (1998),
 surface brightness and integrated magnitudes where applicable, 
a preliminary classification (star, galaxy, etc.), profile shape, 
information about close neighbors, etc.  For objects with SDSS spectroscopy,
links to the spectra will be added.

The spectra are made available as one-dimensional, wavelength and 
flux calibrated tables with 4096 pixels each.  Redshifts are being
derived from emission lines and absorption lines separately, and 
both are listed.  Furthermore, a list of lines, a link to the 
photometric and astrometric data, and an explanation of why the
target was selected for spectroscopy are provided.  

The SDSS data products remain proprietary within the SDSS for approximately 
1.5 years in the beginning.  The proprietary period for new data
will be reduced as the survey progresses.  The data
will be released in several installments to the astronomical
community through dedicated ftp and WWW servers and on CD-ROM.  Mirror
sites for data distribution are currently being set up at STScI, in
Japan, and probably also in Germany.  An easy-to-use web-browser-based
query tool that will link image and catalog data is currently being developed.  

The early data release of the SDSS commissioning data (5\% of the imaging
survey)
is planned for June 6, 2001 during the summer meeting of the American 
Astronomical Society.  The first main release is scheduled for
January 1, 2003 (15\% of the photometric data), followed by releases 
in 2004 (47\% and 68\%), 2005 (88\%), and 2006 (everything).  Since
there is some time delay between imaging and spectroscopy, the fraction
of released spectra will initially lag behind the amount of publicly available 
photometry data.

\section{SDSS Science Results}

Already during its commissioning phase and in its first year the SDSS
has produced a wide range of scientific results as reflected by numerous
conference presentations and refereed papers.  
A good measure of the success
of a project are its scientific findings and their presentation in the
refereed literature (both in terms of papers and citations).  It is 
too early to review the number of citations of SDSS papers since the
majority of them were only published in 2000 and 2001, but we can 
consider the  number of refereed publications:  More than 30 papers
have been published in or submitted to refereed journals since 1999. 
In 1999 three SDSS science papers were published.  In 2000 the number
rose to 12. 
Within the same period, six technical papers were published in or
submitted to peer-reviewed journals.  
In the following sections, we will
try to give an overview of the various science areas addressed by
the SDSS and highlight the science results obtained so far.

\subsection{Large-Scale Structure}

A survey like the SDSS is tailored to study large-scale structure.  Indeed
this is the main purpose of the SDSS.  A homogeneous set of high-quality
imaging data and redshifts will enable us to measure galaxy clustering over
an unprecedented scale, as a function of redshift, and of galaxy type.  
The mass density of the universe,
$\Omega$, can be determined from anisotropies in the three-dimensional
spatial galaxy distribution in the SDSS.
The angular correlation function can be 
determined.  Large-scale peculiar velocity flows can be detected.  
Though the presently existing data cover only a small fraction of the final
survey area, they already reveal strong clustering in the distribution of
bright red galaxies
out to redshifts of $z=0.45$ and little evidence for clustering among
quasars out to $z=2.5$.  

\subsection{Clusters of Galaxies}

Clusters of galaxies play a vital role as probes of large-scale structure
and of cosmological mass density as a function of redshift.
The individual constituents of galaxy clusters are useful for determining
the galaxy luminosity function as a function of environment and redshift,
for studying the morphology-density relation, and 
evolutionary probes.

The SDSS data are being used to compile a comprehensive, uniform galaxy 
cluster catalog through objective, repeatable, automated techniques.
A variety of cluster finding algorithms are being used for this
purpose including Voronoi tessellation, matched filter techniques, 
and other enhancement techniques (e.g., Goto et al. 2001).  The 
usage of the very homogeneous image data combined with multi-color
information aids
in this enterprise.  The current searches concentrate on redshifts of
$0 < z < 0.6$.  Follow-up studies include spectroscopic confirmation of
members and correlation with other data bases such as ROSAT's RASS.

\subsection{Weak Lensing}

Gravitational lensing is caused by masses distributed near the line of
sight to distant objects.  The statistical effect of galaxy-to-galaxy 
lensing leads to small distortions in the shapes of background galaxies
through foreground galaxies (weak lensing).  In order to measure
these effects one needs to make assumptions about the intrinsic shapes of
the background galaxies, which should not be distorted by, e.g., 
star formation or interactions, or by alignment effects
within galaxy clusters.  Contaminating effects of this kind are best
minimized statistically by sampling over large areas under good seeing
conditions.  Redshift information helps to determine whether background
galaxies are spatially correlated.
The amount of the shape distortion, or shear, is a direct measure of the
mass of the foreground lens.  

Fischer et al. (2000) used SDSS commissioning data to measure galaxy-to-galaxy
weak lensing and detected the shear signal out to radii of $600''$ with 
high statistical significance.  They find that the dark halos of luminous
foreground galaxies extend to 260 $h^{-1}$ kpc.  The shear produced by
ellipticals is stronger than the shear produced by spiral galaxies, consistent
with the ellipticals being more than two times more massive.  

Sheldon et al. (2001) selected galaxy clusters found in both the SDSS and 
in RASS data and detected highly significant shear
consistent with an isothermal density profile, demonstrating that ensemble
cluster masses can be measured from SDSS imaging data.  They estimate that
the SDSS will ultimately contain more than 1000 galaxy clusters with 
RASS cross identifications, which will allow one to measure the 
correlation between lensing mass and cluster X-ray luminosity as well as 
the ratio of luminous to dark matter.

\subsection{New Quasars in the SDSS}

One of the primary SDSS science goals is to detect new quasars 
and to obtain spectra of $10^5$ quasars.  
The pre-selection of quasar candidates is done in color space and allows
one to detect, in principle, quasars with redshifts of $0 < z < 7$.

A comparison of quasars detected and spectroscopically confirmed by
the SDSS with catalogs of known quasars reveals that the new SDSS quasars
roughly triple the number of quasars within a given region in the
sky (Richards et al.\ 2001a; Fig.\ 2).  
During its commissioning phase the SDSS already established several
records in quasar detection:  More than 150 quasars with $z>3.5$
were found, including nine of the ten highest-redshift quasars known, and six
quasars with $z>5$.  The most distant object known to date was discovered
in the SDSS: a quasar with $z=5.8$ (Fan et al.\ 2000; Fig.\ 3).  
Interestingly, this quasar shows already metal emission lines and evidence
for a highly ionized universe only $\sim 1$ Gyr after the Big Bang.
The derived black hole mass of this object is $3 \times 10^9$ from
Eddington arguments (absolute magnitude of $-27.2$ at a rest-frame
wavelength of 1450 \AA ).

\begin{figure}[ht]    
\centerline{\vbox{
\psfig{figure=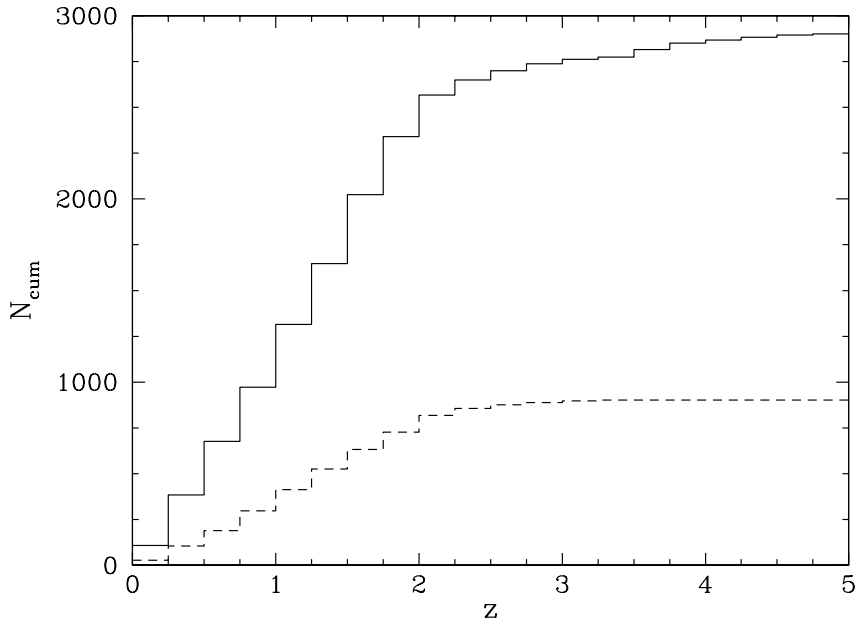,width=12cm,angle=270}
}}
\vspace{-0.3cm}
\caption{ \label{fig_QSOs}
{
Cumulative number distribution (N$_{\rm cum}$) of quasars within a $\sim 530$ 
square degrees equatorial stripe as a function of redshift z.  The 
dashed line denotes all quasars previously listed in the pre-SDSS 
literature.  The solid line comprises all these and the new quasars 
detected by the SDSS.  The SDSS detections increased the number of
known quasars in this region by 200\%.  Adapted from Richards et al.\
(2001a). 
}
}
\vspace{-0.1cm}
\end{figure}

In SDSS data the most distant radio-loud quasar was discovered.
The number of BAL quasars was significantly increased (e.g., 
Anderson et al.\ 2001).  Furthermore, a sample of intrinsically reddened
quasars was found.  Follow-up studies are in progress.
The SDSS led to the discovery of high-redshift quasar pairs, whose 
de-projected separation is estimated to be $< 1$ kpc (e.g., Schneider et
al.\ 2000).  The detection of such pairs provides evidence of clustering
at $z>3.5$.

\begin{figure}[ht]    
\centerline{\vbox{
\psfig{figure=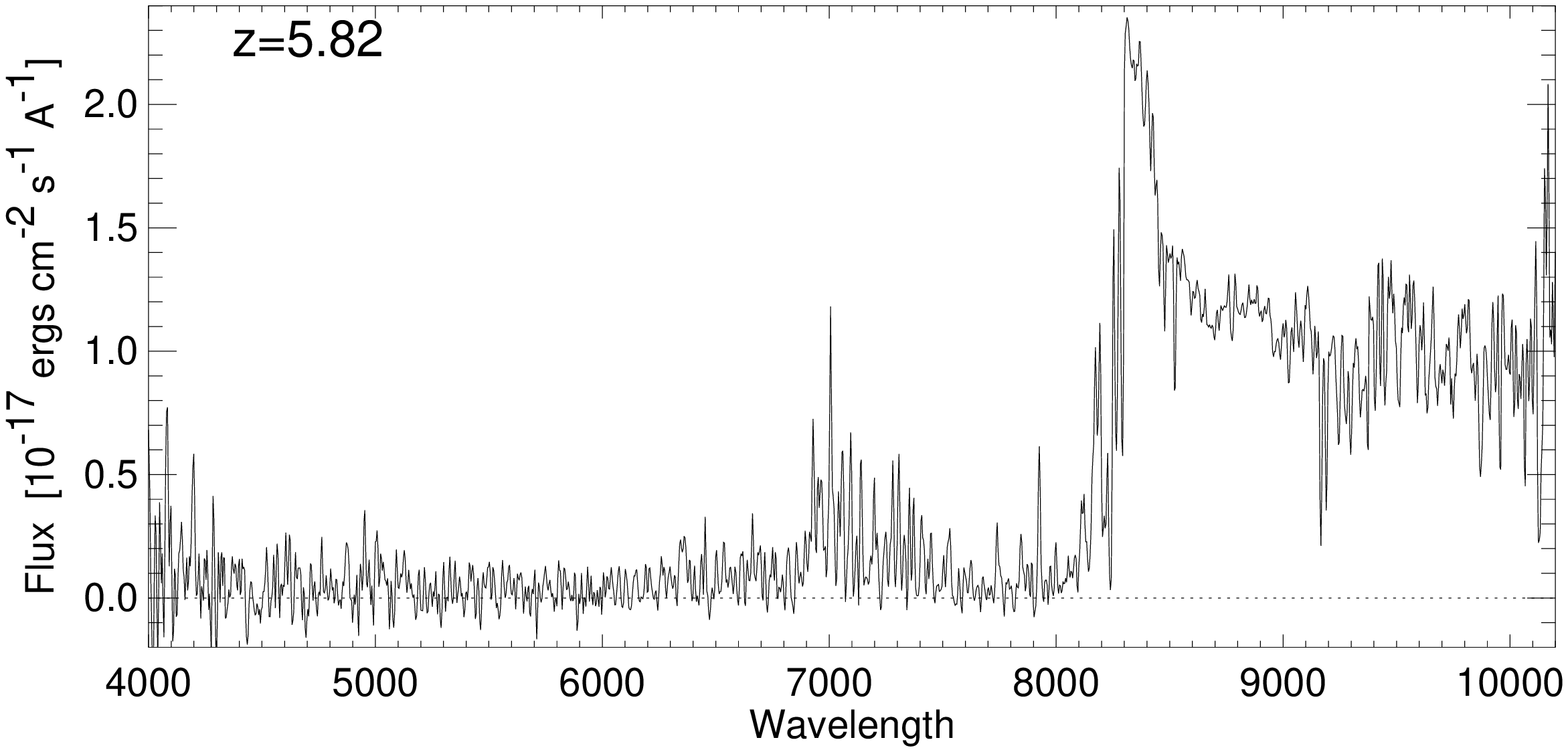,width=12cm,angle=90}
}}
\vspace{-0.3cm}
\caption{ \label{fig_QSOz}
{
Spectrum of the most distant quasar known so far ($z = 5.8$; Fan et al.\
2000).  This spectrum was obtained with Keck II/ESI.
The flux level of the Lyman $\alpha$ forest does not reach zero.
This lack of the Gunn-Peterson trough indicates that at this redshift
the universe is already highly ionized.  
Image credit: Richard White, Space Telescope Science Institute.
}
}
\vspace{-0.4cm}
\end{figure}

Ultimately the SDSS data will allow us to study the quasar luminosity function
as a function of redshift at high spectral resolution and in unprecedented
detail.  Furthermore, the much improved quasar census will contribute
significantly to studies of large-scale structure.
The high-redshift quasar luminosity function seems to be
considerably shallower than
at low-redshift (Fan et al.\ 2001).  Preliminary results for the quasar
spatial density indicate a pronounced decrease at $z>3.5$ and, in 
combination with 2dF results, a peak somewhere between redshifts of 2 and 
3 (Fan et al.\ 2001). 

The five-filter SDSS photometry is uniquely suited to determine photometric
redshifts since the SDSS filters cover a wide wavelength range with little
overlap.  The four colors constructed from adjacent filters
mimic a low-resolution objective-prism survey with $R\sim 4$ 
ranging from $\sim 3000$ \AA\ to $\sim$ 10,000 \AA\ (Richards et al.\
2001b).  These colors show a strong correlation with redshift, and little
dispersion at a given redshift.  Richards et
al.\ (2001b) find that 70\% of the photometrically determined redshifts are
correct to within $\Delta z$ = 0.2 for $0 < z < 5$ down to a magnitude
of $g' \le 21$ mag.   Hence SDSS photometry has the potential of adding 
$10^6$ quasar candidates to the $10^5$ quasars for which SDSS will obtain
spectra.  

Vanden Berk et al. (2001) combined more than 2200 SDSS quasar spectra
to produce composite spectra covering a rest wavelength range of
800\,\AA\ -- 8555 \AA\ with 2 \AA\ resolution and a S/N $<$ 300.  These 
spectra contain more than 80 emission lines.  There is  a clear trend of
increased contributions from young and intermediate-age populations in
the underlying host galaxies at decreasing redshift.  
These composite spectra are the highest-resolution, largest wavelength-range
spectra existing to date and will be extremely useful for a variety 
of applications including template fitting and quasar continuum determinations.

\subsection{Galaxy Studies With the SDSS}

The SDSS is obviously a perfect tool for galaxy studies.  The combination
of galaxy photometry, surface profiles, structural parameters, spectra,
and redshifts makes it possible to address a wide range of topics.
The properties of galaxies in voids vs.\ groups vs.\ clusters can be 
studied with excellent statistics. Galaxy morphology, the 
morphology--density relation, and galaxy luminosity functions
can be investigated as a function of environment, type, and redshift.
Galaxy profiles can be used for disk/bulge decomposition, studies of 
lopsidedness, indications of interactions, etc. 
The homogeneity of the imaging data makes the SDSS ideally suited for
the search for low-surface-brightness galaxies and dwarf galaxies.
The galaxy spectra yield a wealth of information useful for deriving
star formation rates, emission and/or absorption line abundances, 
and the characteristics of the dominant underlying populations.

Blanton et al. (2001) determined the luminosity function for 11,275 galaxies
with SDSS redshifts and absolute magnitudes in the range of approximately 
$-23$ mag $< M_r < -16$ mag.  The resulting luminosity density exceeds that
of the Las Campanas Redshift Survey by a factor of two owing to the use of
Petrosian magnitudes, which include a larger fraction of the total galaxy
flux than uncorrected isophotal magnitudes.  Luminosity functions derived
by earlier surveys can, however, be reproduced when a similar isophotal
magnitude criterion is used.  Blanton et al. find the Schechter function
to produce good fits.  The low-luminosity slope is similar in all five
SDSS bandpasses, while the slope $\alpha$ is shown to be sensitive to 
galaxy surface brightness, color, and morphology.  Luminous galaxies 
generally have higher surface brightness, redder colors, and are more
concentrated than less luminous galaxies, consistent with earlier results.
Blanton et al. also find a strong magnitude -- surface-brightness relation.

\subsection{Galactic Structure Studies With the SDSS}

A wide-area, multi-color, homogeneous imaging is an ideal tool for
Galactic structure studies.  Star counts within the vast area probed
by the SDSS will lead to improved determinations of the structure and 
position-dependent scale heights of
the Galactic thick and thin disk and the Galactic halo.  

The commissioning data have been used
with great success to search for substructure in the Galactic halo.
These studies led to the discovery of overdensities in ``A-colored
stars'' identified as blue horizontal branch stars and blue stragglers
(Yanny et al.\ 2000), and RR Lyrae (Ivezi\'{c} et al.\ 2000).
Ibata et al.\ (2001) showed that this substructure is consistent with
the expected location of the stellar tidal stream torn off the Sagittarius
dwarf spheroidal galaxy that is currently being accreted by the Milky Way.
As the area surveyed by the SDSS grows one expects to trace the tidal
streams over larger regions, allowing us to constrain not only the
dynamical history of Sagittarius, but also the mass distribution of the
outer Milky Way.  

Searches for stellar substructure can be carried out in
arbitrary regions in the hope of serendipitous discoveries. 
Alternatively, one can target 
the surroundings of objects that may produce such 
substructure.  Both star clusters and nearby dwarf galaxies fall in the
latter category, since they may suffer tidal disruption through the
Milky Way.  Numerical simulations predict that possibly as many as half 
of the present-day Galactic globulars will not survive for another
Hubble time.   Prime candidates for tidal disruption are diffuse,
low concentration 
clusters.  

\begin{figure}[ht]    
\centerline{\vbox{
\psfig{figure=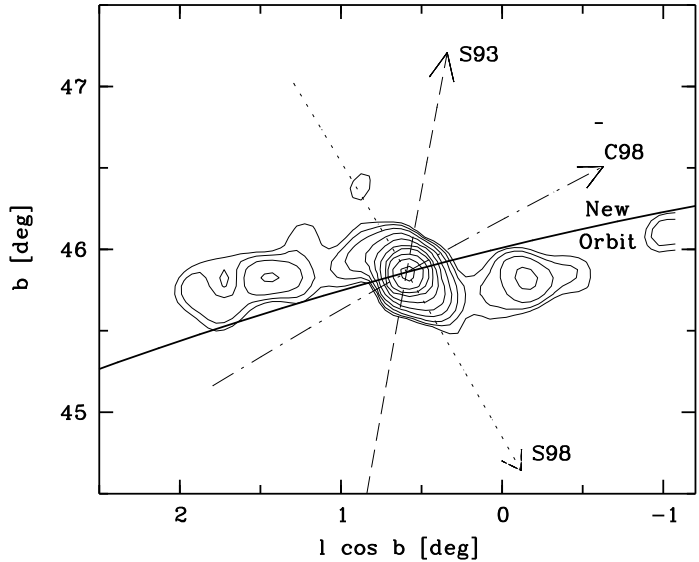,width=12cm,angle=270}
}}
\vspace{-0.2cm}
\caption{ \label{fig_pal5}
{Contour plot of the surface density of color-magnitude-selected
candidate member stars of the Galactic globular cluster Pal 5
in galactic 
coordinates. The overlaid arrows represent 
different orbital paths of the 
cluster according to different determinations of its absolute
proper motion.  The solid line presents our improved estimate of the orbit
(Odenkirchen et al.\ 2001).
}
}
\vspace{-0.4cm}
\end{figure}

Among the first star clusters observed during the 
SDSS commissioning phase was the sparse, distant halo globular cluster
Palomar 5.  We searched for stellar overdensities in color-magnitude space
around Pal 5 and discovered two well-defined, symmetric tidal tails  
around Pal 5 that subtend an arc of at least $2.6\degr$ on the sky 
(Odenkirchen et al.\ 2001).  The tails show an S-shaped structure near
Pal 5 and exhibit
two density clumps at equal distances to the cluster center, both in accordance
with expectations from N-body simulations.  The stars in the tidal tails
make up 34\% of the cluster's stellar population, showing that the cluster
is suffering heavy mass loss and may be completely torn apart after its
next disk passages.  The orientation of the tails 
allows us to constrain the cluster's orbit. 

Studies of the structural parameters and a search for potential tidal
tails around dwarf spheroidals around the Milky Way and other Galactic
globular clusters are in progress.  

\subsection{Special Types of Stars in the SDSS}

Stars observed by the SDSS provide an invaluable resource in their
own right.  While only stars that appear to be peculiar in some way 
(such as through unusual colors) are scheduled for SDSS spectroscopic
follow-up, the five-passband photometric database allows one to perform
a preliminary classification of special types of stars based on their
distinctive colors alone.  
Stars that stand out due to their colors include white dwarfs, cataclysmic
variables, hot subdwarfs, carbon stars, M, L, T, and brown dwarfs.

\begin{figure}[ht]    
\centerline{\vbox{
\psfig{figure=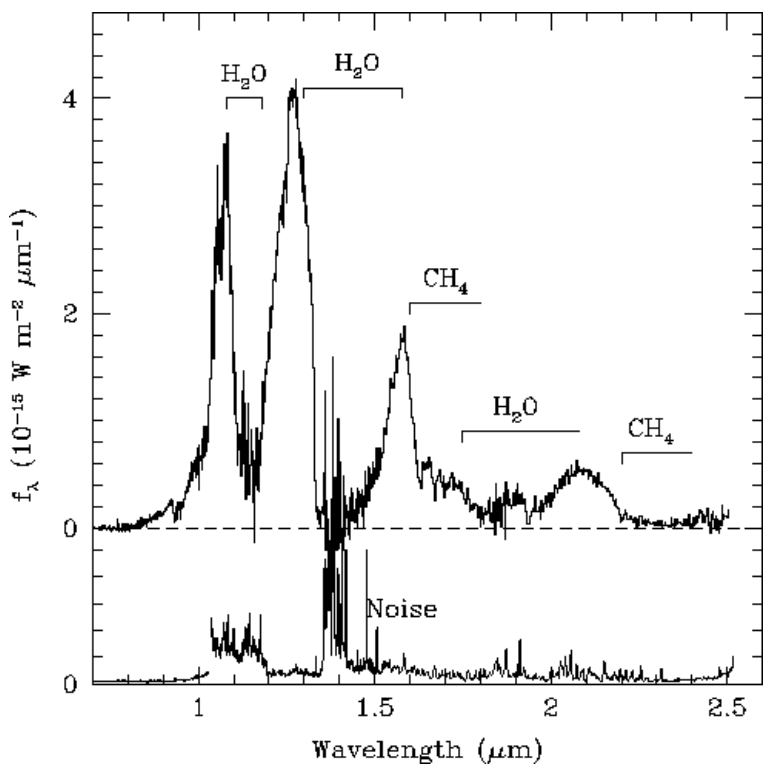,width=10cm,height=10cm}
}}
\caption{ \label{fig_methane}
{Spectrum of 
the first methane dwarf detected by the SDSS (Strauss et al.\ 1999).  
This brown dwarf is a free-floater
in the field at a distance of $\sim 10$ pc.  
The strong methane and water bands in the spectrum show that
it is a member of the rare class of T dwarfs.  The spectrum was obtained 
with UKIRT.  Image credit: SDSS Collaboration.
}
}
\vspace{-0.1cm}
\end{figure}

The SDSS is rapidly increasing the white dwarf census and may as much as 
quintuple it.  The individual white dwarfs cover a wide range of types
including cataclysmic variables (22 new objects discovered at the time
of writing), magnetic stars, or very cool white dwarfs,
making detailed studies of the white dwarf luminosity function possible.
Some of the SDSS white dwarfs are constituents of white dwarf --
red dwarf pairs.  Hot white dwarfs are often central stars of planetary
nebulae.  

The coolest white dwarf known to date was recently
discovered by the SDSS (Harris et al.\ 2001).  Very cool white dwarfs
are important since they belong to the oldest objects in the universe.  
In order to detect these faint objects they need to be fairly nearby. 
The proper motion of the recently discovered,
very cool (3000 K -- 4000 K) white dwarf
indicates that it belongs to the Galactic disk rather than the halo.

The SDSS is also improving the Galactic C star census, particularly for
faint, high-latitude C stars.  The latter are believed to be distant
halo objects,  which would make them excellent
tracers of the kinematics of the Galactic halo, and of the Milky Way
potential.  So far 36 new C stars have been discovered,
half of which are dwarf C stars.  Dwarf C stars are a type of star
that was only recently recognized.  Owing to their faintness only nearby
(within $\sim 100$ kpc) objects of this class have been detected so far.
Preliminary findings indicate that these objects are the dominant type
of C star.  Together with very cool white dwarfs they may contribute
significantly to the baryonic mass of galaxies.  Due to their proximity
proper motions can be determined for dwarf C stars, which together with
their radial velocities constrains their motion.  

Cool late-type stars are another type of stars whose numbers are 
significantly increased by the SDSS.  Many of these objects lie in the
same SDSS color range as high-redshift quasar candidates.  The 
combination with 2MASS data is helping to remove ambiguities and to
identify promising targets for spectroscopic follow-up.  The SDSS
has already discovered more than 100 new late M dwarfs, about 50 
new L dwarfs, and more than 15 T dwarfs.  The spectral classes L and
T were added to the spectral sequence only within the past three years
(Kirkpatrick et al.\ 1999).  
Low-mass stars and brown dwarfs cooler than late M dwarfs ($\sim 1400
< T_{\rm eff} < 2000$ K) are classified as L dwarfs.  In L dwarfs
TiO and VO absorption decreases, while the H$_2$O bands increase
with decreasing temperature. The T dwarf
range begins at temperatures cooler than 1300 K.  T dwarfs may be the
link between stars and gaseous planets and are sufficiently cool to 
allow methane to form in their atmospheres.   T dwarfs are characterized
by pronounced H$_2$O and CH$_4$ absorption (Leggett et al.\ 2000).
Very cool, red stars identified by the SDSS were recently found to
establish the missing spectral transition link between the L and T classes and
helped to provide a full spectral sequence for these new classes from
late M to late T dwarfs (Leggett et al.\ 2000).  

\begin{figure}[ht]    
\centerline{\vbox{
\psfig{figure=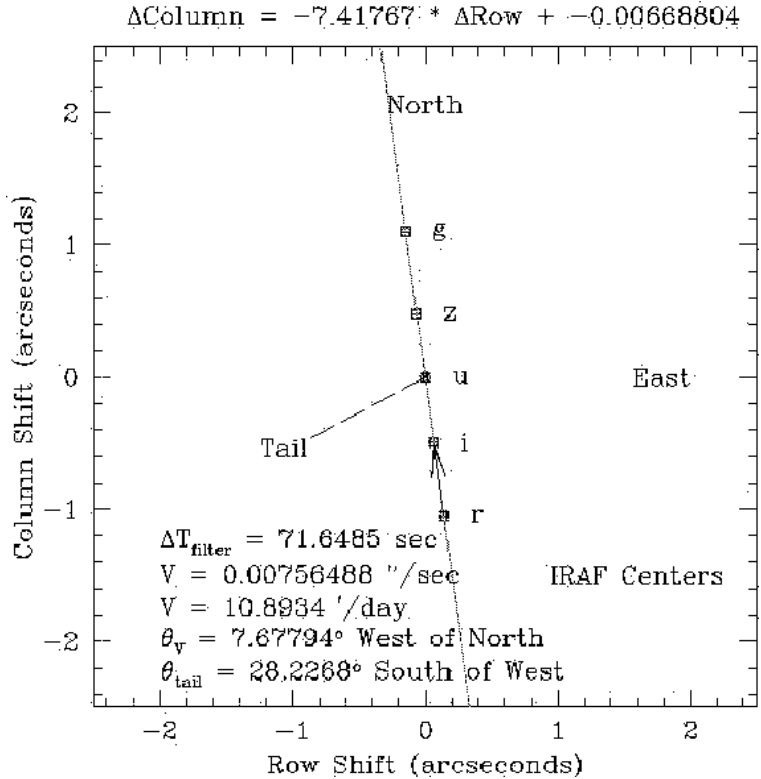,width=10cm,angle=270}
}}
\vspace{-0.3cm}
\caption{ \label{fig_comet}
{An illustration of the SDSS's ability to detect moving solar system
objects: 
The motion of comet Dalcanton across the sky as it was observed by the 
successive filters of the SDSS imaging camera during a scan.  
Image credit: Julianne Dalcanton. 
}
}
\end{figure}

Two field methane brown dwarfs, or T dwarfs, have been discovered in the SDSS
so far (Strauss et al.\ 1999; Tsvetanov et al.\ 2000; Fig.\ 5).
They both lie at distances of approximately 10 pc, have estimated
masses of 0.015 -- 0.06 $M_{\odot}$, and ages of 0.3 to 5 Gyr.  
More of these rare free-floaters were detected in DENIS, 
2MASS, and in proper motion
surveys.  Their number may amount to a few hundred across the entire sky.

\subsection{Solar System Science With the SDSS}

Nearby moving objects such as asteroids and comets show clearly detected motion
within an SDSS scan.  The SDSS software picks out these objects automatically.
The SDSS already found quite a few asteroids  and
its first comet (Fig.\ 6).   This comet is one of the few objects believed
to have come from the inner Oort Cloud and has its perihelion near the
orbit of Jupiter (Dalcanton et al.\ 1999).  More distant objects such as
Kuiper Belt objects are too far away to have detectable motion
in a single SDSS scan.

\section{The Calar Alto Key Project for SDSS Follow-up}

While the SDSS is still in its first year of regular operations, it
has already contributed a wide range of impressive discoveries.  It 
is also obvious that various areas of SDSS-based science need to be
complemented by observations that the SDSS itself cannot provide.  
Infrared imaging or spectroscopy is essential for a variety of studies
from quasars to low-mass stars.  Stellar spectroscopy is intentionally
neglected by the SDSS since cosmological problems are among its primary
science goals.  Certain types of SDSS follow-up studies require 
long-slit spectroscopy, others deeper imaging than a drift-scan survey
can provide, yet again others need observations at very high angular
resolution.  

The SDSS scientific returns can be maximized by adding targeted 
targeted follow-up observations, at wavelength
and spectral set-ups not covered by the SDSS itself, and by providing
these capabilities at a guaranteed-time basis.

The instrumentation at Calar Alto provides a
unique opportunity for such targeted follow-up observations 
for a variety of different programs.
Of the above requirements, Calar Alto can provide infrared imaging and
spectroscopy, optical longslit spectroscopy, and deep imaging.
The director of MPIA therefore decided to set aside a certain fraction
of observing time for a key project for SDSS follow-up studies at
Calar Alto.  This project is in some ways a successor of the previous
MPIA key project, CADIS.   
Having fixed amounts of observing time
allocated to this project throughout each semester during the five-year
lifetime of the survey will allow sufficient
flexibility to accommodate
target-of-opportunity observations as well as long-term planning of
follow-up observations for large SDSS science projects,
thus optimizing the scientific returns.
We anticipate to use 30 -- 45 nights per semester for this
key program, which is expected to run until 2005.
As an in-kind contribution to the SDSS the MPIA is committed to
making part of this observing time available to the SDSS collaboration.

The first semester of key project observations began in
January 2001 with a total number of 44 nights.  These nights include
bright, grey, and dark time at the 2.2-m and 3.5-m telescopes at
Calar Alto.  We are currently using four instruments:  MAGIC for 
infrared imaging and spectroscopy, CAFOS for imaging and low-resolution
optical spectroscopy, MOSCA for optical multi-object spectroscopy,
and TWIN for optical spectroscopy at SDSS resolution and wavelength
coverage, or at higher resolution. 

Time is allocated on a monthly basis in order to be able to respond
quickly.  Incoming proposals are reviewed by the SDSS group at MPIA.
The observations are carried out in service mode at Calar Alto.
Service mode is particularly well suited for the SDSS follow-up 
projects as it allows to accommodate requests for observations in
photometric conditions, good seeing, etc., to be carried out with more
flexibility.  

At MPIA we are currently pursuing five main SDSS projects in 
collaboration with other SDSS astronomers:\vspace{-0.20cm}
\begin{itemize}\itemsep=-0.3pt\parsep=0pt\parskip=0pt
\item Source identification of quasar and low-mass star
candidates, 
\item H$\alpha$ rotation curves of spiral galaxies, 
\item Low-surface-brightness galaxies,
\item Galactic structure and tidal streams,  
\item Template spectra for SDSS population synthesis.
\end{itemize}
In addition, there are programs led by other members in the collaboration.
In the following, we give a brief
overview of SDSS science currently pursued at Calar Alto. 

\subsection{Follow-up on SDSS Quasar and Low-Mass Star Candidates}

At redshifts of 3 and higher quasars become increasingly red due to intervening
absorption-line systems and are well separated from the bulk of
other stellar or galaxy contaminants except for very cool,
low-mass stars such as L and T dwarfs (Fan et al.\ 1999).  Optical
spectroscopy or infrared imaging help to distinguish quasar candidates
not targeted by the SDSS due to, e.g., faintness.  
At MPIA we concentrate on the search for new high-redshift ($z>4.5$)
quasars through multi-color near-infrared
observations of candidates pre-selected through optical SDSS photometry.
In part,
these observations aim at obtaining a complete sample of QSOs with
optical and infrared colors.
Furthermore, we follow up on $z$-band-only detections through additional
$J$-band imaging.  The resulting ($z-J$) colors enable us to separate
low-mass stars (mostly L dwarfs) from QSOs and to select targets for
follow-up spectroscopy with 8-m to 10-m class telescopes. 

SDSS spectroscopy is not usually obtained for science targets brighter
than $r' \sim 14.5$ mag.  Therefore, 
another quasar project executed at Calar Alto and other observatories
aims at obtaining redshifts for quasars brighter than this magnitude
limit.

Calar Alto spectroscopic follow-up is also being carried out for a sample
of T Tauri stars and late-type dwarfs
selected by their SDSS colors.  We hope to ultimately
establish complete
magnitude-limited samples of low-mass stars, an important precondition
in deriving deep stellar and substellar luminosity functions.  
This project will also reveal whether the majority of the elusive T dwarfs exist
as part of binary systems or in isolation in the field, confirm additional
ancient, very metal-poor dwarfs, help to 
verify candidate T Tauri stars and thus improving
their census, and provide a basis for further follow-up
studies with large telescopes.

\subsection{Galaxy Kinematics}

SDSS spectra are of limited use for deriving internal galaxy kinematics
since the spectra are obtained through fibers with a fixed circular diameter
of $3''$.
We use long-slit spectroscopy of an SDSS-selected galaxy sample spanning
a a wide range of luminosity and color in order to obtain 
rotation curves and stellar velocity dispersions.  SDSS provides distances,
luminosities, colors, and morphological parameters for these
galaxies.  The longslit data will provide 
dynamical parameters --- the depth and profile of
galaxy potential wells --- that can be incorporated
into principal component analyses  of the galaxy
distribution.  The data will also yield accurate estimates of the
true scatter in bivariate ``distance indicator''
relations, most notably the Tully-Fisher relation, as a function of 
luminosity and wavelength.
By combining
the correlations between dynamical and photometric
properties obtained from our sample with the
distribution of photometric properties derived from
the full SDSS redshift survey, we can estimate the
distribution of galaxy-mass potential well depths,
a fundamental prediction of galaxy formation theories.
This data set may also be useful for many other
investigations, including searches for new
distance-indicator relations, studies of the
distribution of galaxy angular momenta, and
identification of dynamically peculiar systems.

\subsection{Low-Surface-Brightness Galaxies and Dwarf Galaxies}

This project follows up on low-surface-brightness (LSB) and dwarf galaxies 
that we detect in the SDSS images based on adaptive filters
and other techniques.  The homogeneity, area coverage, and
depth of the SDSS makes it uniquely suited for improving the 
census of these faint objects.  With many of the LSB and dwarf galaxies
too faint to be targeted by SDSS spectroscopy, our
spectroscopic follow-up studies allow us to determine distances,
abundances, kinematics, and to some extent even star formation histories. 
This will vastly increase the available data on low-mass galaxies
and enable us to carry out a thorough study of galaxy evolution and
impacting factors (such as environment) with unprecedented comprehensiveness.
 
\subsection{Galactic Structure and Tidal Streams}

The SDSS data quality far exceeds the capabilities
of photographic all-sky surveys, which are less
deep, have poorer resolution, suffer from large-scale inhomogeneities
and center-to-edge plate variations, and are limited in area coverage.
This makes it an excellent tool for Galactic structure studies as 
described earlier.  Still, the SDSS is limited in depth.  Also, it
does not obtain stellar spectra on a regular basis.   

Calar Alto follow-up on tidal streams comprises both deep imaging and 
spectroscopy.  Deep imaging is used to
obtain luminosity functions in streams and the parent object, and to
study mass segregation.  Spectroscopic follow-up is used both to
establish kinematic membership of stars in streams and to measure the
velocity dispersion, which yields crucial constraints for dynamical
modelling.

\subsection{Calibration and Modelling}

The interpretation of both SDSS photometry and spectroscopy requires
a thorough understanding of the SDSS calibration.  Currently the SDSS
calibration is still considered preliminary, subject to additional
improvements prior to the first large public data release.
This affects the absolute calibration of the SDSS photometry and the
flux calibration of the spectroscopy.  E.g., a still unanswered
question is how SDSS $3''$ fibre spectra of point sources
or extended sources compare to regular longslit spectra of the same
sources.  Another potential concern is how well the combination of
SDSS spectra obtained in the red and in the blue channel works.

A more general topic is the interpretation of spectroscopic indices
measurable in galaxies, in particular Balmer line ratios as
age indicators and Lick indices.  Population synthesis models may
yield different ages even for simple stellar populations than does
detailed color-magnitude diagram modelling of their
resolved stellar populations.  This is a particularly severe problem
that becomes difficult to assess in galaxies where only the integrated
light can be measured.  Age and metallicity indicators like the Lick indices
may be subject to errors owing to the impact of, e.g.,
horizontal branch morphology variations or mixed-age populations
rather than single-age populations.

We are working on establishing a spectroscopic template database at
SDSS resolution.  Our targets include both stars with well-established
metallicities and star clusters with known ages, reddenings,
distances, and abundances.
This effort is being carried out at Calar Alto and other observatories
and will be vital for the interpretation of SDSS spectra and SDSS-specific
population synthesis models.
 
\section{Outlook}

The SDSS is mapping one quarter of the sky through multi-color imaging
and spectroscopy at unprecedented area coverage and homogeneity.  The
reduced and calibrated data will become available to the entire astronomical
community over the next five years. They will undoubtedly spawn a wide range of 
discoveries ranging from cosmology to star formation.  This unique vast
database will change the way we do astronomy and provides a virtual
observatory in itself.  

There are efforts to carry out southern large-sky surveys.  The 2dF 
QSO Redshift Survey (2QZ) will obtain redshifts for more than 25,000
quasars.  ESO is
building the VLT Survey Telescope (VST), a 2.5-m telescope for optical
wide-field imaging.  The British Visible \& Infrared 
Survey Telescope for Astronomy (VISTA) will provide both optical and 
infrared imaging. VISTA will reach approximately five times
deeper than the SDSS in the northern hemisphere and will surpass DENIS and
2MASS in depth as well.   

The SDSS and the ongoing/planned 
southern surveys are complemented by similarly large 
databases in other wavelength ranges such as 2MASS, FIRST, HIPASS/HIJASS, 
RASS, etc.  The combination of multi-wavelength fluxes, velocity and 
distance information may lead to discoveries that we can not yet envision.

\subsection*{Acknowledgements}

The Sloan Digital Sky Survey (SDSS) is a joint project of The University 
of Chicago, Fermilab, the Institute for Advanced Study, the Japan 
Participation Group, The Johns Hopkins University, the Max Planck Institute 
for Astronomy (MPIA), the Max Planck Institute for Astrophysics
(MPA), New Mexico State University, Princeton University, the United States 
Naval Observatory, and the University of Washington. Apache Point 
Observatory, site of the SDSS telescopes, is operated by the Astrophysical 
Research Consortium (ARC). 

Funding for the project has been provided by the Alfred P. Sloan Foundation, 
the SDSS member institutions, the National Aeronautics and Space 
Administration, the National Science Foundation, the U.S. Department of 
Energy, Monbusho, and the Max Planck Society. The SDSS Web site is
                              {\tt http://www.sdss.org/}. 

\vspace{0.7cm}
\noindent
{\large{\bf References}}
{\small

\bref
Anderson, S.F., Fan, X., Richards, G.T., Schneider, D.P., Strauss, M.A.,
et al. 2001, AJ, submitted

\bref
Blanton, M.R., Dalcanton, J.D., Eisenstein, D., Loveday, J., Strauss, M.A.,
et al. 2001, AJ, 121, 2358

\bref
Dalcanton, J., Kent, S., Okamura, S., Williams, G.V., Tichy, M., et al.
1999, IAU Circ., 7194, 1

\bref
Fan, X., Knapp, G.R., Strauss, M.A., Gunn, J.E., Lupton, R.H., et al. 2000,
AJ, 119, 928

\bref
Fan, X., White, R.L., Davis, M., Becker, R.H., Strauss, M.A.,
et al. 2000, AJ, 120, 1167

\bref
Fan, X., Strauss, M.A., Schneider, D.P., Gunn, J.E., Lupton, R.H., et al.
2001, AJ, 121, 54

\bref
Fischer, P., McKay, T.A., Sheldon, E., Connolly, A., Stebbins, A., et al.
2000, AJ, 120, 1198

\bref
Fukugita, M., Ichikawa, T., Gunn, J.E., Doi, M., Shimasaku, K., \&
Schneider, D.P. 1996, AJ, 111, 1748

\bref
Goto, T., Sekiguchi, M., Kim, R.S.J., Bahcall, N.A., Annis, J., et al.
2001, AJ, submitted

\bref
Gunn, J.E., Carr, M., Rockosi, C., Sekiguchi, M., Elms, B.B., et al.
1998, AJ, 116, 3040

\bref
Harris, H.C., Hansen, B.M.S., Liebert, J., Vanden Berk, D.E., Anderson, S.F.,
et al. 2001, ApJ, 549, L109

\bref
Ibata, R., Irwin, M., Lewis, G.F., \& Stolte, A. 2001, ApJ, 547, L133

\bref
Ivezi\'{c}, \v{Z}, Goldston, J., Finlator, K., Knapp, G.R., Yanny, B.,
et al. 2000, AJ, 120, 963 

\bref
Kirkpatrick, J.D., Reid, I.N., Liebert, J., Cutri, R.M., Nelson, B.,
et al. 1999, ApJ, 519, 802

\bref
Leggett, S.K., Geballe, T.R., Fan, X., Schneider, D.P., Gunn, J.E., 
et al. 2000, ApJ, 536, L35

\bref
Lupton, R.H., Gunn, J.E., \& Szalay, A.S. 1999, AJ, 118, 1406

\bref
Odenkirchen, M., Grebel, E.K., Rockosi, C.M., Dehnen, W., Ibata, R., et al.
2001, ApJ, 548, L165

\bref 
Richards, G.T., Fan, X., Schneider, D.P., Vanden Berk, D.E., Strauss, M.A.,
et al.\ 2001a, AJ, 121, 2308

\bref
Richards, G.T., Weinstein, M.A., Schneider, D.P., Fan, X., Strauss, M.A., 
et al.\ 2001b, AJ, submitted

\bref
Schlegel, D.J., Finkbeiner, D.P., \& Davis, M. 1998, ApJ, 500, 525

\bref
Schneider, D.P., Fan, X., Strauss, M.A., Gunn, J.E., Richards, G.T., et al.\
2000, AJ, 120, 2183

\bref
Sheldon, E.S., Annis, J., B\"ohringer, H., Fischer, P., Frieman, J.A., 
et al. 2001, ApJ, in press

\bref
Strauss, M.A., Fan, X., Gunn, J.E., Leggett, S.K., Geballe, T.R., et al.
1999, ApJ, 522, L61

\bref
Tsvetanov, Z.I., Golimowski,D.A., Zheng, W., Geballe, T.R., Leggett, S.K., 
et al. 2000, ApJ, 531, L61

\bref
Vanden Berk, D.E., Richards, G.T., Bauer, A., Strauss, M.A., Schneider, D.P.,
et al. 2001, AJ, submitted

\bref
Yanny, B., Newberg, H.J., Kent, S., Laurent-Muehleisen, S.A., Pier, J.R.,
et al. 2000, ApJ, 540, 825

\bref
York, D.G., Adelman, J.E., Anderson, S.F., Anderson, J., Annis, J., et al.
2000, AJ, 120, 1579

}

\vfill

\end{document}